\documentclass[fleqn,10pt]{wlscirep}

\usepackage{comment}

\title{Balance of thrones: a network study on \textit{Game of Thrones}}

\author[1,2,3,*]{Dianbo Liu}
\author[2,4,5,6,7]{Luca Albergante}

\affil[1]{Computer Science and Artificial Intelligence Lab, Massachusetts Institute of Technology, Cambridge,MA,USA}
\affil[2]{School of Life Sciences, University of Dundee, Dow Street, Dundee, UK}
\affil[3]{The Broad Institute of MIT and Harvard,Cambridge,MA,USA}
\affil[4]{Computational and Systems Biology of Cancer, Institut Curie, Paris, France}
\affil[5]{PSL Research University, Paris, France}
\affil[6]{Institut National de la Santé et de la Recherche Médicale U900, Paris, France}
\affil[7]{Mines ParisTech, Paris, France}

\affil[*]{Corresponds to: dianbo@mit.edu}

\begin{abstract}
TV dramas constitute an important part of the entertainment industry, with popular shows attracting millions of viewers and resulting in significant revenues. Finding a way to explore formally the social dynamics underpinning these show has therefore important implications, as it would allow us not only to understand which features are most likely to be associated with the popularity of a show, but also to explore the extent to which such fictional world have social interactions comparable with the real world. To begin tackling this question, we employed network analysis  to systematically and quantitatively explore how the interactions between noble houses of the fantasy drama TV series \textit{Game of Thrones} change as the show progresses. Our analysis discloses the invisible threads that connected different houses and shows how tension across the houses, as measure via structural balance, changes over time. To boost the impact of our analysis, we further extended our analysis to explore how different network features correlate with viewers engagement and appreciation of different episodes. This allowed us to derive an hierarchy of features that are associated with the audience response. All in all, our work show how network models may be able to capture social relations present in complex artificial worlds, thus providing a way to qualitatively model social interactions among fictional characters, hence allowing a minimal formal description of the unfolding of stories that can be instrumental in managing complex narratives. 
\end{abstract}
\begin{document}
\maketitle
\pagebreak

\flushbottom

%
%
\thispagestyle{empty}

\section*{Introduction}

Network modeling is a powerful tool to represent interactions between different agents (usually called nodes), which can represent either physical or abstract entities\cite{newman2003structure,newman2011structure}. Once, a network has been constructed, it can be used to formally explore the different features observed, and hence to quantitatively assess the modes of interaction among the agents. Social sciences have a long history of using networks to explore human interactions\cite{barabasi2002linked}, with the classical experiment of Milgram, paving the way to popular concepts like social hubs and \textit{six degrees of separation} between individuals.

While many theoretical approaches focused on studying the structures of networks with only one type of interaction (e.g., network of friends), in a few cases authors considered more rich models, where each interaction is associated with one of more interaction types. One of such theories is behind the concept of \textit{structural balance}\cite{heider1946attitudes,cartwright1956structural,harary1980simple,kunegis2010spectral}. In this theory, the nodes represent individuals that can form positive or negative relationships with the others. Using a theoretical approach it is also possible to show that the system will converge toward two ideal states: either all the individual display a positive attitude or the individuals forms two camps displaying a negative attitude towards each other\cite{antal2006social}. 

Network analysis has been extensively applied in the field of social sciences\cite{waumans2015topology}, but its application to the study of dynamical changes of a social network over time are more limited, partly due to the complexity of controlling confounding factors such as uncontrollable external event. We therefore sought to explore the potential of network modeling to explore the isolated fictional world where the popular American fantasy drama TV series \textit{Game of Thrones} takes place. The series has been adapted from George Martin's novel \textit{A Song of Ice and Fire} \cite{martin2016song} and has multiple interconnected plot lines and we decided to focus on the story arc unfolding on the fictional continent "Westeros" where noble houses of the "Seven Kingdoms" fight for the "Iron Throne" or independence. In employing network analytics to explore the structures and dynamics of the popular TV show, we assessed whether network theory could be used to quantify the changes associated with the evolution of the story and to gain insight into the reaction of the viewers to such changes\cite{facchetti2011computing,marvel2011continuous,antal2005dynamics,beveridge2016network}.

\section*{Results}

\subsection{Constructing a social network of houses}

Instead of focusing on individual characters of the drama, we decided to model the interactions among large \textit{meta-entities}, such as major noble houses, armies, religious groups and other groups with significant political influences. Networks of relationships among the meta-entities were built for each episode in the first six seasons of the \textit{Game of Thrones} TV series, in such a way to explore the changes episode by episode. To provide a more focused interpretation, the analysis was restricted to interactions that took places in 'Westeros', a frictional isolated continent where most of the actions are localized. Hence, events that took places beyond 'the wall' and to the east of 'the Narrow Sea' were not taken into consideration. 

\begin{figure}
\centering
\includegraphics[width=0.7\textwidth]{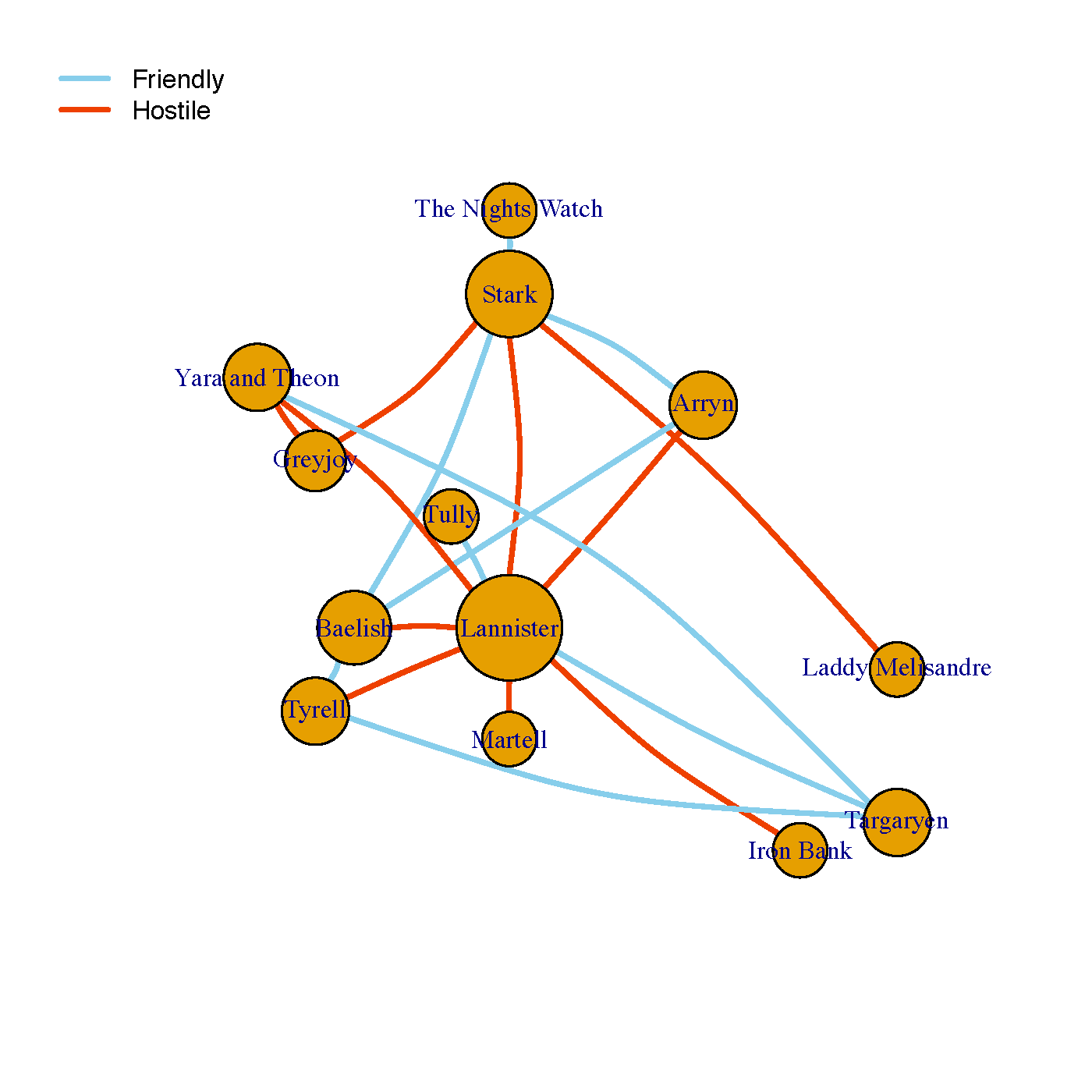}
\caption{\label{fig1} Network of relationships among meta-entities in \textit{Game of Thrones}. This network models the last episode of season six. Only activities on the 'Westeros' continent were taken into consideration.}
\end{figure}

The complicated relationships among the meta-entities were abstracted and represented using a signed networks. Each node indicates a meta-entity and each edge represents a political relationship between two entities (Figure \ref{fig1}). A edge was marked as either positive (colored in blue in the figures) --- if the two entities were on friendly terms during the episode --- or negative (colored in red in the figures) --- if the relationship was hostile. Only political relationship involving actual interactions were taken into account. If two entities were indifferent to each other or did not have real interactions, no edge was drawn between the two nodes. 

\subsection{Large-scale properties of the networks}
\begin{figure}
\centering
\includegraphics[width=0.7\textwidth]{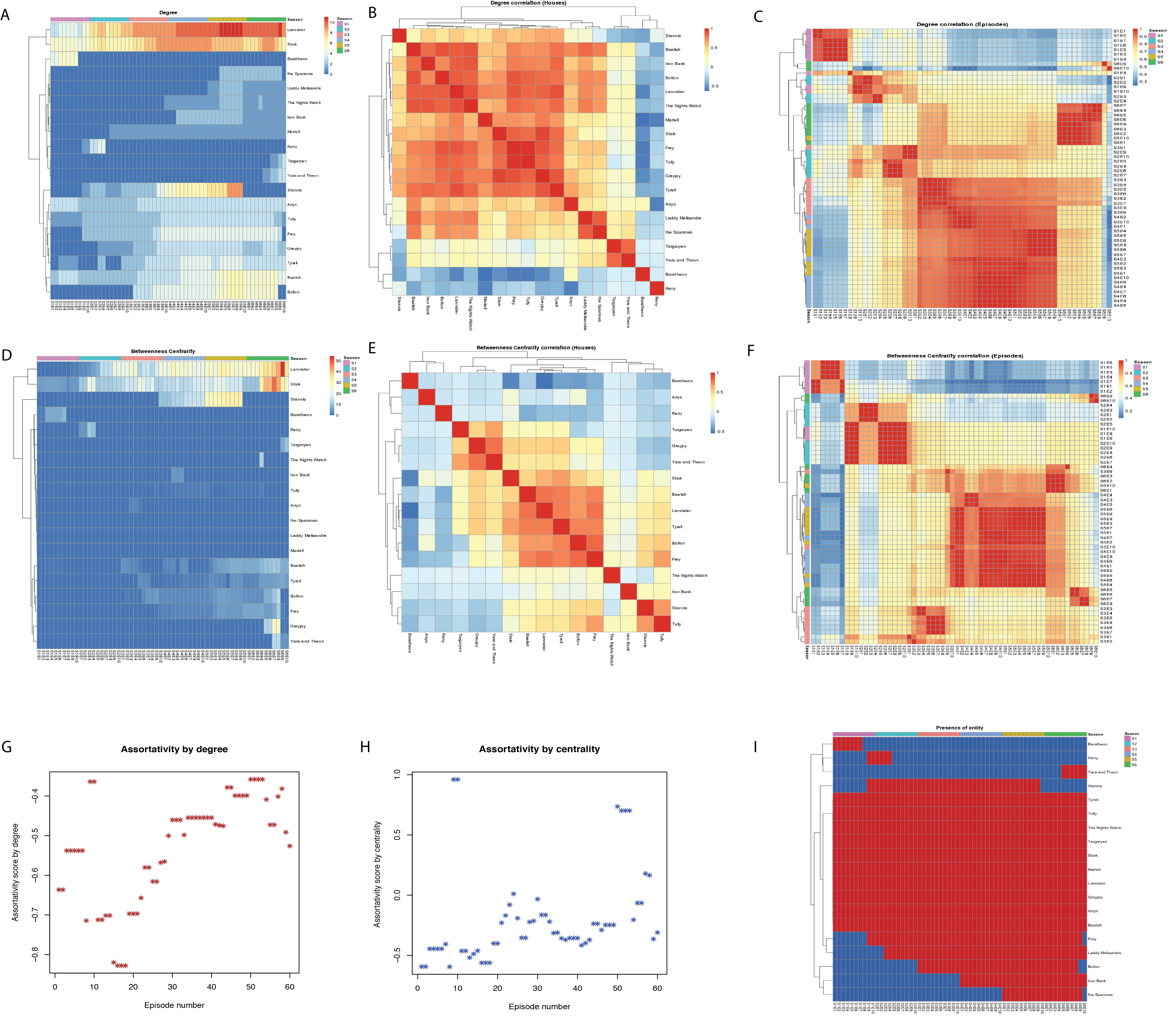}
\caption{\label{fig2} Large Scale Properties of the networks. (A) Node degree (connections to other node) of each noble houses/political entity in season 1-6 of \textit{Game of Thrones}. (B) Correlation between the node degrees by house across the the episodes. (C) Correlation between the node degrees by episode across houses. (D) Betweenness centrality of each houses/political entity. (E) Correlation of betweenness centrality by episode across the the houses. (F) Correlation of betweenness centrality by house across the the episodes. (G) Changes of assortativity by degree over the first six seasons of the TV series. (H) Changes of assortativity by betweenness over the first six seasons of the TV series. (I) Presence(red) and absence(blue) of different entities over time. }
\end{figure}

As a first step in exploring structural patterns in the relationships among the meta-entities, large-scale network properties were considered, starting with degree-based measures. The degree of a node in a network is the number of direct edges connecting this note to the others and is related with the importance of a node in the network\cite{newman2010networks}, as node with larger degree interact with more entities. When the degree of all of the nodes was derived for the network associated with each episode (Figure \ref{fig2}A), we found that the political entities taking part in most of the story lines, such as 'House of Stark' and 'House of Lannister', are associated with nodes possessing a very high degrees throughout a large part of the episodes considered. This is intuitively understandable as a highly active house, or army, is more likely to have interactions with others.

To further this analysis, and find potentially less obvious connections between the houses, we considered the correlation of degrees among different entities(Figure \ref{fig2}B) and across episodes(Figure \ref{fig2}C). Degree correlation among houses suggested similarity in dynamics of importances in the story. First of all, it is interesting to note a predominance of positive correlations in both representations. From the meta-entities points of view, this supports the ideas that most meta-entities tend to adjust the number of interaction in a coordinated way, suggesting the idea that the proportion of total number of interactions remain somewhat stable over the episodes. While from the episode point of view, this indicates that, across episodes, the number of interactions distribute in a somewhat proportional way.

Looking at the dendrograms accompanying the plots it is also possible to observe a number of strongly defined clusters. One example is the strong correlation between the degrees of the 'House of Frey' and 'House of Tully'. In the show, these two houses were geologically located close to each other and had complicated political interactions, this justifying the finding. Another interesting observation is the high correlation between the "the Sparrows" and "Laddy Melisandre", both of which were representatives of religious group and were heavily involved in political conflicts, which suggested that some common patterns of how religions influenced on the progresses of the story. 

Looking at the correlation episode-wise, our analysis highlights how the different episodes from the same season tend to be closed together in term of clustering (see row annotations of Figure \ref{fig2}C). Overall, this suggests that different seasons mostly contained episodes with coherent importance groups, as expected from the flow of the story. Looking at the data from a more general point of view, we can also observe cross-season connections. For example, season 3 to 5 were quite similar to each other (see the lower half of the heatmap) while season 1 occupy a rather independent place.       

Network theory employs the notion of centrality to measure the extent to which a node is situated at the \textit{center} of the network and hence is structurally positioned to influence, either directly or indirectly, a large part of the network. One classical measures of centrality is \textit{betweenness centrality}, which measures the number of shortest path passing trough a node \cite{newman2010networks}. According to our analysis, major political entities in the story had high betweenness centrality scores throughout the episodes, as expected. On the other hand political entities that are geologically isolated by 'the narrow sea' or 'the wall' generally had lower betweenness centrality scores, e.g. 'the Night's Watch'(Figure \ref{fig2}D). As we did for the degree, we computed the correlation of the betweenness centrality score between each house and each episode (Figure \ref{fig2}E and \ref{fig2}F). Two medium-size clusters emerged from the analysis with the bigger one containing the 'House of Lannister' and the political entities that had strong interactions with it (Figure \ref{fig2}E). Additionally, most episodes in season 3 to 5 formed again a big cluster, consistent with degree analysis. 

Assortativity measures the tendency of nodes to connect to similar nodes in the network \cite{newman2010networks,newman2002assortative} and can be considered for different properties. For example, Assortativity by degree is the tendency of a node to be directly connected to others with similar degrees and assortativity by betweenness centrality is the tendency of a node to be directly connected to others with similar betweenness centrality. In both scenarios, negative assortative scores were observed in most of the episodes (Figure \ref{fig2}G and Figure\ref{fig2}H), indicating that nodes with similar properties tended not to be connected. Not all the entities were present in all the episodes of the first six seasons. Presence and absence of the entities were summarized in figure \ref{fig2}I.

\subsection{Changes of relationships correlated with number of votes and ratings from viewers}
\begin{figure}
\centering
\includegraphics[height=0.4\textwidth]{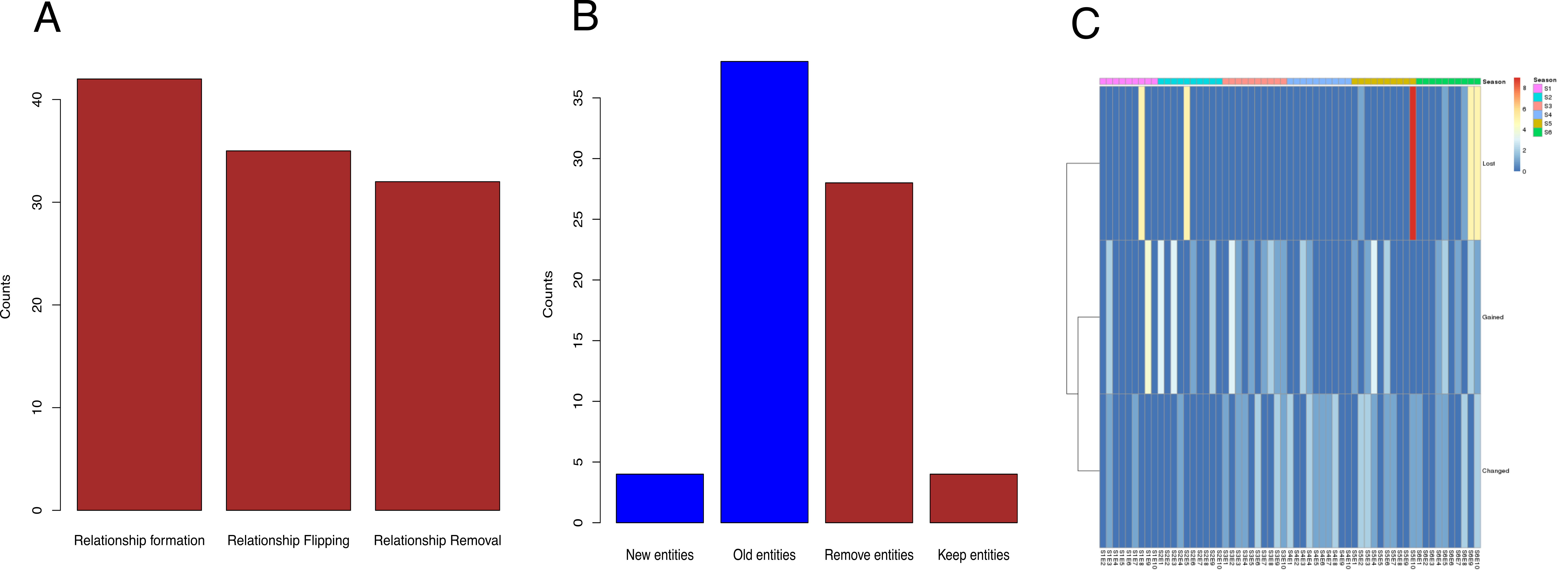}
\caption{\label{fig3} Changes of edges in the network overtime.(A) Formation of new edges was more frequent than flipping signs and removal of edges. (B) Most new edges were formed between existing nodes (blue). Most edge removals were coupled with elimination of political entities (brown). (C) Changes of edges in each episode. }
\end{figure}

Figures \ref{fig2}A and \ref{fig2}D suggested that substantial changes were present in the topological structure of the network as the story progressed, and hence that the interactions among the meta-entities changed perceivably across episodes. To further investigate this issue and better interpret the results of our analysis, we classified network changes from one episode to the next into three categories: 1) 'relationship establishment' (a new edges was added, indicating the establishment of a new relationship), 2) 'relationship flipping' (the sign of a edges was changed, indicating that allies became enemies or enemies became allies), and 3) 'relationship disruption' (an edge was removed, indicating a previously present relationship was removed due to either two entities ceasing to interact or to the removal of one, or both, of the interacting entities). The establishment of a new relationship was the most frequent change, followed by flipping and disruption (\ref{fig3}A and C). It is also worth mentioning that while most new relationships were formed between existing political entities (87.5\% of the changes), relationship disruption was mostly due to the elimination of entities (Figure \ref{fig3} 3B).

\subsection{Dynamics of structural balance of the network were different from real social networks}
Although a network representation is a powerful way to describe the structure of a system of interacting entities, the actual dynamics of the system depends on many specificities that are abstracted away when a network is produced. Nonetheless, it is sometime possible to associate certain types of dynamics with particular network structures \cite{AlonBook, ECY:ECY1973543638, BQS, LongChain}. In the context of social interaction, the theory of structural balance has been used to formally explore how positive (e.g., friendship) and negative (e.g., hate) interactions can produce situations that are more or less balanced due to the psychological stress of the interacting agents. Structural balance is grounded in social psychology theories pioneered by Heider in 1940s \cite{heider1946attitudes} and has been previously used to study the evolution of human interactions in different contexts \cite{antal2006social}.

\begin{figure}
\centering
\includegraphics[width=0.7\textwidth]{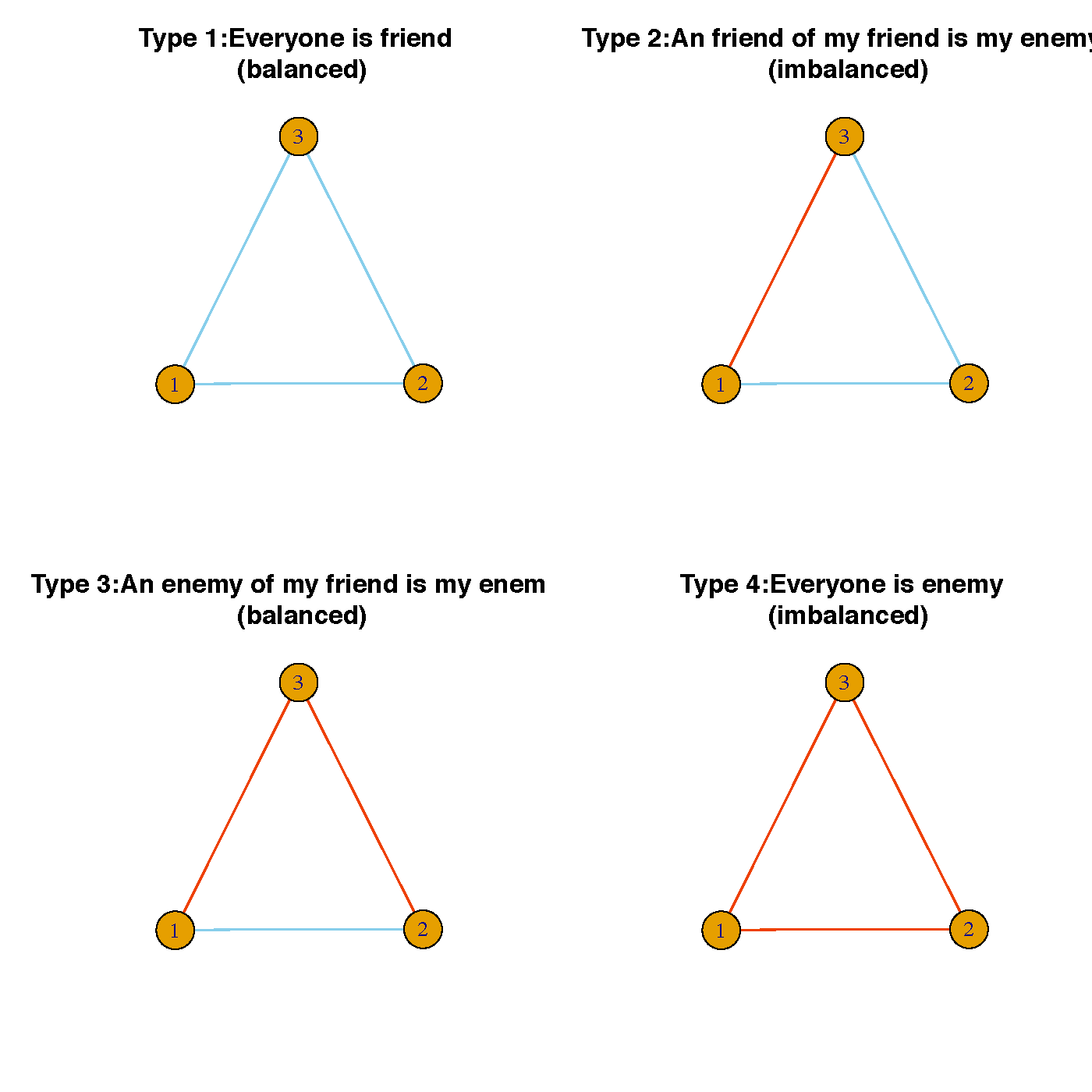}
\caption{\label{fig4}Structural balance. The complex network of \textit{Game of Thrones} can be divided into a large number of constituent triads. If everyone is allied or if any two members are allied against a common enemy, the triad is balanced. If everyone is hostile to others or a single entity is allied to two entities that are enemies with each other, then the triad is imbalanced. A network is balanced if each constituent triad is balanced\cite{antal2006social} }
\end{figure}

Structural balance is based on classifying triads (i.e., a groups of three entities that interact with each other), into balanced and unbalanced depending on the sign of the interactions. More formally, given two meta-entities of ‘Game of Thrones’ $i$ and $j$, we associate a value to the edge $S_{ij}$ connecting the two entities so that $S_{ij}=1$ if they are allies and $S_{ij}=-1$ if they are enemy. Then, a triad consisting of three meta-entities $i$ , $j$ and $k$ is balanced if $S_{ij} \cdot S_{jk} \cdot S_{ik}=1$ and imbalanced if $S_{ij} \cdot S_{jk} \cdot S_{ik}=-1$. The idea of balanced is based on the intuitive notion that if all of the members of the triad are allied (all edge are positive) or two members of the triad are allied against the third one (exactly one edge is positive), then the situation is balanced and there is not pressure for the members of the triad to change mode of interaction. In all the other cases, at least one of the member is under a pressure that will produce an imbalance and hence a likely change \cite{antal2006social}. A network is said to be balanced if each constituent triad is balanced (Figure \ref{fig4}).  

\begin{figure}
\centering
\includegraphics[width=0.7\textwidth]{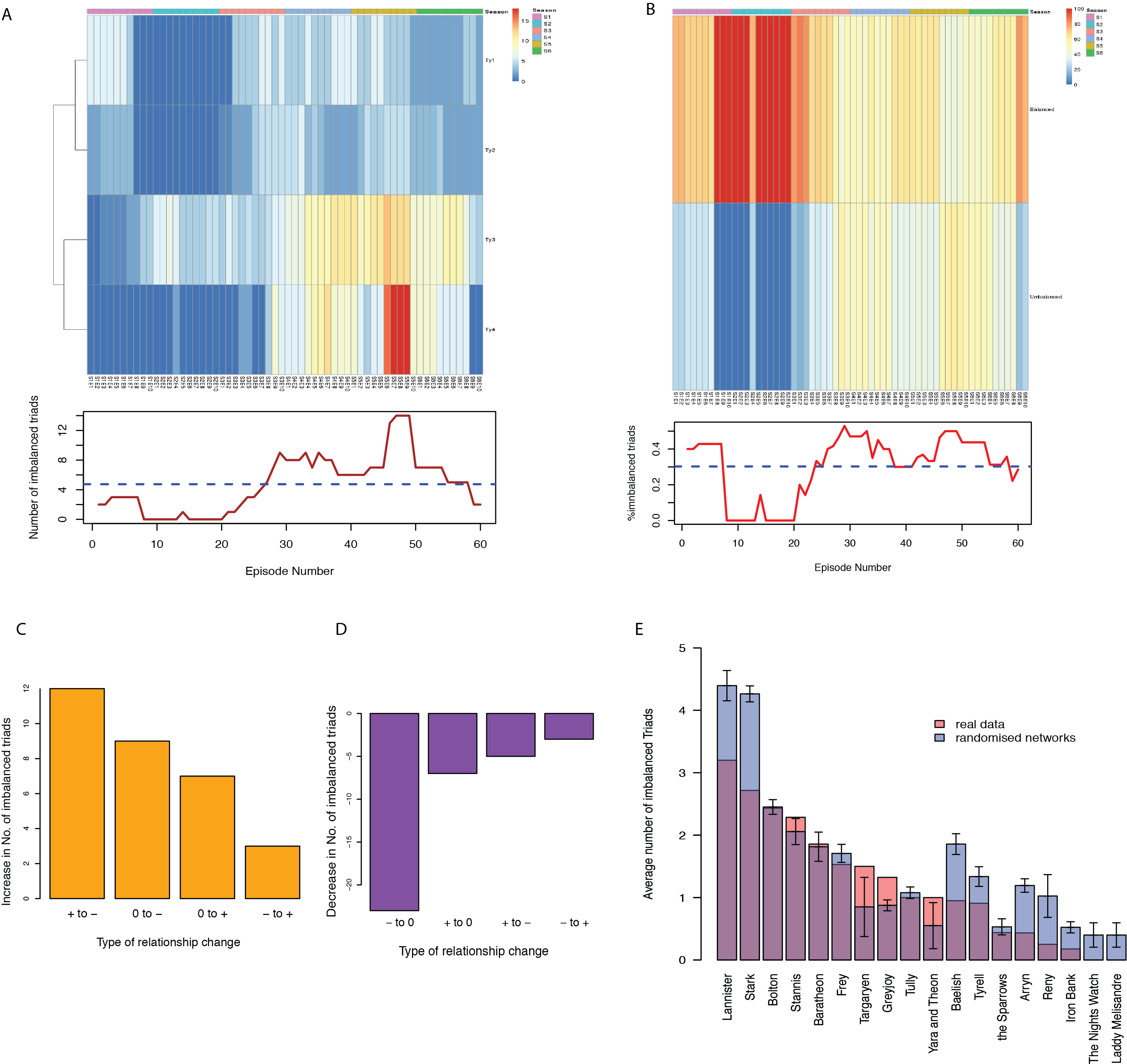}
\caption{\label{fig5} Structural balance of the network. (A) Number of different types of triads (upper panel) and number of imbalanced triads over time (lower panel).(B) Percentage of balanced (upper panel) and imbalanced triads (upper and lower panel) over time. (C) Contribution of different types of edge changes to the increase in the number of imbalanced triads. "+" stands for positive edge , '-' for negative 'edge', and '0' represents the absence of relationship. (D) Contribution of different types of edge changes to the decrease in the number of imbalanced triads. (E) Average number of imbalanced triads per episode for each entity. Red bars indicate the real data extracted from \textit{'Game of Thrones'} and blue bars indicate the expected number of imbalanced triads from randomised networks. Error bars indicate standard deviation.}
\end{figure}

It was suggested that in social and political networks, relationships change to minimize the number of imbalanced triads\cite{antal2006social}. However, such trend seems to be absent  in the network of \textit{Game of Thrones}. The number and percentage of imbalanced triads through the first 6 seasons seemed to fluctuate over time with an average value of 4.75 which corresponded to about 30 percentages of all triads (Figure \ref{fig5}A and \ref{fig5}B). This phenomenon can be interpreted as a strategy by the author, directors and screenwriters to keep the story engaging, since in a balanced network, no political entity would be pressured to change and the story would potentially stall. Interestingly, a temporary tendency towards balance was observed at the beginning of the show when two dominating alliances initiated a total war against each other ('the War of Five Kings'). However this balance was broken as with the story progresses (or perhaps was broken to \textit{allow} the story to progress). 

The change of friendly relationship to hostile or the introduction of hostile interaction between previous non-interacting entities were the most common expedients to introduce imbalance in the networks (Figure \ref{fig5}B), while the level of imbalance was reduced mainly by eliminating hostile interactions via the disappearance of political entities (Figure \ref{fig5}C).

The average numbers of imbalanced triads each house was contributing to over the evolution of the story were also considered (Figure \ref{fig5}D). Hub of imbalance, and thus thus instability, can be identified by our analysis. These hubs were associated with houses that are involved in many stories lines of the TV series, such as 'Stark' and 'Lannister'. To better assess this result, we also explored to which extent, the result observed was compatible with randomized networks (i.e. network obtained by shuffling interaction between the entities while maintaining the degree of each entity unvaried). This analysis suggests that most of the entities were taking part in a number of imbalanced triads that is lower than expected by chance alone and hence that the story introduced a controlled mix of balanced and imbalanced triads, in such a way to prevent the system from being overwhelmed by tensions, due to a large number of imbalanced triads, and to avoid creating a completely balanced systems, which would have no pressure to change.

\subsection{Strategy to introduce unpredictability}
\begin{figure}
\centering
\includegraphics[width=0.7\textwidth]{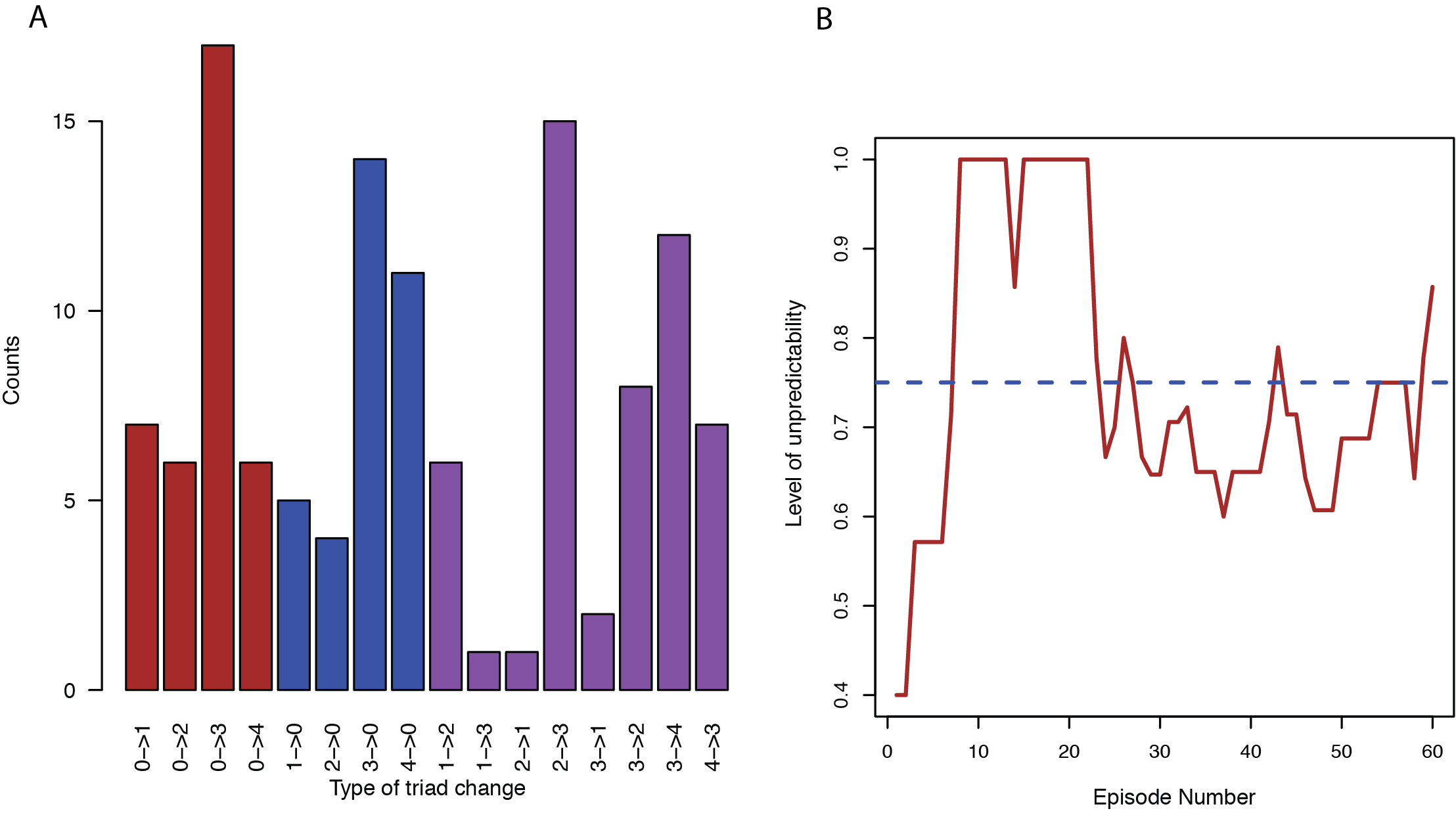}
\caption{\label{fig6}Statistics of dynamics of triads.(A)Summary of all triads changes in the first six seasons. A large percentage of changes involved triads of state 3(B)Changes of levels of unpredictability of the network over time. The level reached 100\% during the "the War of Five Kings" }
\end{figure}

Unpredictable twists in the storyline of a TV show are common expedients to keep viewers engaged and introduce new plot elements. Using our analytical framework, we sought to quantify the extent to which this is happening in \textit{Game of Thrones}. To this end, we classified triads into different types according to types of interaction present (Figure \ref{fig4}). Establishment, elimination and change of type of all triads present in at least one episode of the show are summarized in Figure \ref{fig6}A. Triads of type 3, corresponding to balanced situations where two allies are hostile towards a common enemy, were the most frequently affected by changes, with >47\% of triad formation, >86\% of triad state changes and >40\% of triad disappearance. 
 
One possible explanation for this finding is that triads of type 3 can be transformed into any of the two imbalanced triads by changing the type of only one edge. Therefore, this type of triads provides a balance starting point that can subsequently be used to introduce imbalance in a more \textit{soft} way. Hence type 3 triads can be used to start a story arc and potentially to let it die. Indeed, we found that a large number of type 3 triads get destroyed,  which was a good proxy for the end of a subplot. Moreover, we can see that many triads changed from, and into, type 3, indicating that the interaction mode supported by this triad was a common intermediate step (Figure \ref{fig6}A).

Since triads of type 2 and 3 can be both changed into two different types, we reasoned that they could be used as proxy of the unpredictability of the story (as in both cases two choices are possible, in addition to the removal of the triad). To better quantify the level of unpredictability as previously defined, we introduced an unpredictability score defined as
\[
U=\frac{T_{3}+T_{2}}{\sum_{i}T_{i}}
\]
where $T_{i}$ is the number of triads of state $i$. $U$ takes values between 0 and 1, indicating, respectively, low and high unpredictability. Our analysis suggests that unpredictability peaks in season 2 during "the War of Five Kings" which were two major wars happened in \textit{Game of Thrones}.

\subsection{Correlation between network properties and viewers' responses}

Up to this point, a number of network topological and dynamic features of the
\textit{Game of Thrones} network have been explored. One question that could be asked is whether these network properties could be used to dissect viewers' responses to the TV show and therefore estimate how interesting the story is. 

Viewers' votes and ratings were both taken into consideration in this study. Number of votes and viewers' ratings are organized for each episode (Figure \ref{fig7} a and c). Spearman partial correlation was conducted to understand potential correlation between votes or ratings with a variety of network properties. Number of nodes, number of edges , number of edge changes , percentage of imbalanced triads , number of triads and assortativity were included into this analysis. 
P values of association of votes with different network properties are 0.66, 0.99,0.79 ,0.29, 0.68, 0.0017 respectively. The correlation coefficients are -0.059,0.00041 ,-0.036 ,0.14 ,-0.056 ,0.41 (Figure \ref{fig7} b).  This indicated that network assortativity has a significant positive correlation with number of vote for each episode.  P values of association of ratings with the network properties are 0.064, 0.42,0.88 ,0.98,0.93, 0.14 respectively. The correlation coefficients are -0.25 ,0.11, 0.020, -0.0033, 0.011,0.20 (Figure \ref{fig7} d).  This shows that larger numbers of nodes of the interaction networks likely correlates with lower ratings.

\section{Discussion}
Structures and dynamics of complex systems or stories consisting of many entities and relationships are usually difficult to understand  and simplified model can be very effective in exploring the key properties of such systems. In this article we followed this philosophy and employed network modeling to quantify structural changes over the story arc of the \textit{Game of Thrones} TV show by capturing both topological and dynamics changes in a systematic way. 

The relationship network was abstracted at the level of noble houses or political entities to rely on a high level representation that limits the uncertainty associated with single characters. Degree distribution and centrality of the network were shown to be an efficient way to identify important political entities. The fact that the major houses of the story, such as "House of Lannister" and "House of Stark", had much higher numbers of connections than others is consistent with findings in other networks and provides a formal way to asses their importance in the unfolding of the story. The negative assortativity observed in the network has structural similarities to natural systems such as marine food web, with minor entities gravitating around major players. Such structure is also quite dissimilar to human systems such as co-authorship networks. Overall, this seems to suggest a somewhat fractal organization of the story that allows the authors to concentrate on story arcs tightly associated with major players, while allowing a large number of minor plots entangled with the mayor story. Hence allowing a multilevel story-telling that can be engaging for different viewer types and may contribute to boost the interest in the show.

In addition to static topologies, the dynamics of the network was tracked through the 60 episodes composing the first six seasons of the show. The positive correlation identified between number of edges changes per episode and viewer's votes suggested that dramatic changes in relationships between houses can make the series more attractive to audiences, perhaps as a consequence of the reorganization in the balance of powers.

Furthermore, since the edges of the network can be associated with an positive or negative interaction between the entities, the theory of structural balance can be employed to explore how different types of three-body interactions contribute to the evolution of the story. This dynamic analysis allows us to explore the preferred story-telling structures in a formally sound way, highlighting the different importances of the various interaction modes and suggesting effective ways to unfold story in a natural, but still engaging, way. To our best knowledge, this is the very first time structural balance theory was used to explore the dynamics of fiction and TV series.

The results from partial correlation analysis suggest that assortivity and number of nodes correlated with viewer's responses. The assortativity coefficient measures the tendency of nodes to connect to similar nodes in the network. Its correlation with number of votes possibly reflects clustering of similar houses in sense of interaction patterns raise population of the show among audiences. On the other hand, the negative correlation between number of nodes and ratings potentially suggest that an over complex story is not favored by the viewers. 

Writing a novel or a TV show is a complex process that requires skills and imagination. Successful authors are able to create stories that stimulate the interest of the readers or viewers, by engaging fictional worlds with their own life. By employing a mathematical approaches to analyze such fictional stories, we were able to show how formal methods can be used to provides new perspectives that complement more classical text analysis. All in all, our works suggest that the methods developed in the context of network theory can complement the tool sets already available to story writers to plan and explore how the dynamics of interaction among single fictional entities contribute to the complex web of relationships that support unforgivable stories.

\section*{Methods}

\subsection{Construction of relationship network}
Each major political entity located on 'Westeros' or possessing strong connections with entities on 'Westeros' was modeled as node in a network. If a house split into more than political entities, they were considered different nodes. One example is the 'House of Baratheon', that split into one army led by Stannis and another army led by Renny. If two nodes have actually interactions with each other, a edge is included between them. All the edges are associated with a color indicating the type of interaction. Friendly relationships were marked as positive (blue) and hostile relationships were marked as negative (red)(Figure \ref{fig1}). If two entities did not interacted directly, no edge was included between the corresponding nodes. For certain edges, where the relationships were complicated, only the effects of real actions were taken into consideration.  

\subsection{Dynamics of the the network}
Changes of nodes and edges were tracked at the resolution of each episode and statuses of nodes and episodes at the end of each episode were compared. Network analysis in this article were conducted using igraph package in R environment (version3.4). 

\subsection{Number of votes and rating from viewers}
The votes and rating from viewers were downloaded from IMdb \url{http://www.imdb.com/title/tt0944947/eprate} on August the $2^\textrm{nd}$, 2017. Both quantities were normalized by dividing over the mean of each season. This was done to account for potential variation in popularity  of the show across seasons.   

\subsection{Network randomization}
To estimate the expected number of imbalanced triad each node was entangled in, signs of edges in each episode were randomly shuffled. A total of 30 different randomized networks were combined to calculate the expected number of imbalanced triads.

\subsection{Data and code availability}
The network data and the code used to write the article are available at \url{https://github.com/kaiyuanmifen/BalanceOfThrones}.

\bibliography{sample}

\section*{Acknowledgements}
We would like to thank Timothy Newman, Rana Al Mamun, and Sam Palmer for interesting discussions and insights.
\section*{Author contributions statement}
DL constructed the networks, provided original concepts, developed the code to perform the analysis, and wrote the manuscript. LA provided original concepts, supervised the work, and wrote the manuscript.

\pagebreak

\end{document}